\documentclass[a4paper]{jpconf}
\usepackage{graphicx}
\usepackage{glas}

\begin{document}

\begin{titlepage}{GLAS-PPE/2015-05}{29$^{\underline{\rm{th}}}$ October 2015}
\title{Enabling Object Storage via shims for Grid Middleware}

\author{Samuel Cadellin Skipsey$^1$, Shaun De Witt$^2$, Alastair Dewhurst$^2$, David Britton$^1$, Gareth Roy$^1$, David Crooks$^1$
\\
$^1$School of Physics and Astronomy, Kelvin Building, University of Glasgow, Glasgow, G12 8QQ \\ $^2$STFC Rutherford Appleton Laboratory, Harwell Oxford, Didcot, Oxfordshire, OX11 0QX}

\begin{abstract}The Object Store model has quickly become the basis of most commercially successful mass storage infrastructure, backing so-called "Cloud" storage such as Amazon S3, but also underlying the implementation of most parallel distributed storage systems. Many of the assumptions in Object Store design are similar, but not identical, to concepts in the design of Grid Storage Elements, although the requirement for "POSIX-like" filesystem structures on top of SEs makes the disjunction seem larger. As modern Object Stores provide many features that most Grid SEs do not (block level striping, parallel access, automatic file repair, etc.), it is of interest to see how easily we can provide interfaces to typical Object Stores via plugins and shims for Grid tools, and how well experiments can adapt their data models to them.
We present evaluation of, and first-deployment experiences with, (for example) Xrootd-Ceph interfaces for direct object-store access, as part of an initiative within GridPP\cite{GridPP} hosted at RAL. Additionally, we discuss the tradeoffs and experience of developing plugins for the currently-popular {\it Ceph} parallel distributed filesystem for the GFAL2 access layer, at Glasgow.

\vspace{0.5cm}
\begin{center}
{\em 21st International Conference on Computing in High Energy and Nuclear Physics (CHEP2015)}\\
{\em Okinawa, Japan}
\end{center}

\end{abstract}
\newpage
\end{titlepage}
\section{Introduction}
Data storage technologies have developed significantly since the dawn of the WLCG, both within and without the community. Within the community, there has been a transition from gridFTP as the {\it de facto} WAN protocol, and various local protocols (rfio, dcap etc.) negotiated by a high level protocol (SRM) and a common Logical File Catalog technology for global replica management; to a brace of WAN protocols (HTTP, xrootd) with lighter negotiation, and a fragmentation of file catalogues across the WLCG experiments. However, the model for representation of storage at a given Storage Element has remained resolutely filesystem like, with SRM and all of the popular SE implementations (CASTOR\cite{CASTOR}, DPM\cite{DPM}, dCache\cite{dCache}) exporting a common namespace with a hierarchical, directory-like, indexing of file names within it.

In the wider world, development has been, perhaps, more rapid and radical, with Object Stores supplanting global file systems as the operative mode of representation (and HTTP the {\it de facto} standard for much of the data transfer to and from these Object Stores). This has been driven by the rise of commercial cloud storage providers (with Amazon S3 at the vanguard), but the technological arguments for Object Storage are also compelling and apply to Grid storage as well.

In particular, the current model for WLCG data management places much of the replica management within the baliwick of the individual experiment infrastructures, with ATLAS' Rucio\cite{Rucio}, CMS' PhEDEx\cite{PheDEX} and LHCb's DIRAC\cite{DIRAC} having overrun the ground originally occupied by the LFC. In particular, Rucio is something of a departure from previous designs; dispensing with a file catalog in favour of algorithmically generated file paths for a given file. The constraint of having to interact with filesystem style storage interfaces significantly complicates Rucio's algorithm - it must guarantee unique names for a given file, whilst also distributing files across directory hierarchies to respect file-per-directory limits and other issues with filesystem scaling. Ironically, this implementation also causes problems for SEs, as the deeply nested directory structures resulting take up much more metadata space in backend databases, and can cause performance issues for global filesystem operations.

\subsection{Object Stores}
Scalability issues due to centralised metadata were one of the core issues that Object Storage was designed to remove\cite{object1,object2}, compared to filesystems. The design of an Object Store explicitly decentralises control and responsibility for data, allowing the separation of metadata and data operations, and the location of data-specific operations close to that data (as a uniquely identified Object). This works well for situations where data is generally written once and read many times, as this decentralisation causes difficulties with maintaining strict state synchronisation across the entire store. 

Common WLCG storage technologies such as DPM and CASTOR attempt to decentralise some operations, locating metadata in a central database and distributing data across multiple file servers. Data access is localised to the fileserver to the extent that TURLs provided by SRM access are usually explicit references to services running on the fileserver with the file, not the central management machine. The global state of an SE is still explicitly synchronised via the central bottleneck, however, as all lookups and handoffs must pass through it.
Many experiments have explored caching TURLs for individual files in order to optimise their performance, thus attempting to achieve object-store-like mechanics by the back door.

We suggest that a hypothetical Rucio-like data management layer backed by Object Store interfaces at SE-like endpoints would be significantly simpler to manage, as Object Storage provides a flat namespace, with objects distinguished by GUID-like hashes. Rucio already maintains an internal GUID for every file object managed by it, so exposing this as a global object name would be trivial. As jobs running on Grid infrastructure need to perform almost no metadata operations, a pure data interface to Tier-2 storage would not impede their performance at all, modulo security considerations.

\subsection{Ceph}
Of particular popularity at present is the distributed Object Store technology called Ceph\cite{ceph}. Ceph's principal innovation is in its CRUSH\cite{sage} algorithm for data distribution amongst constituent filesystems (OSDs), which is designed to minimise the amount of data movement required due to changes in cluster geometry.
For this purpose, the {\it rados} library provides the underlying Object Store based on these principles. Releases of Ceph are assigned common names from the set of common names for cephalopod species, ordered alphabetically; the current release is {\it Hammer}, and the preceding were {\it Giant} and {\it Firefly}.

Ceph also provides higher-order storage abstractions on top of the RADOS Object Store; Rados Block Device interfaces (which resemble a uniform block device), Ceph Object Store (an S3-compatible ``Cloud" object interface) and the POSIX filesystem layer {\it cephfs}. For our purposes, however, the bare Object Store interface is sufficient; removing the higher order abstractions gives us higher potential efficiency.

\subsection{radosstriper}
One caveat with the low-level rados interface is that it provides as simple an Object Store interface as possible. In particular, the rados library provides no data striping functionality for objects - unless the pool is a erasure coded pool (in which case, the object will be coded and striped appropriately), a single object is represented by a single extent of data on a single OSD. For objects on the order of the size of physics data, a few gigabytes, this is very inefficient for both space use and io distribution.
Striping of data is usually provided by the higher level abstractions, in mutually incompatible ways (the RBD backend cannot understand the striping performed by the S3 backend, for example).

The radosstriper\cite{sebastien} library was developed by Sebastien Ponce, originally for use in the CASTOR Ceph backend. It is a part of the {\it Giant} (and later) releases of Ceph. radosstriper provides a lightweight layer over the basic rados interface, providing hooks for selecting the number and size of chunks into which an object will be split. (The chunks are trivially identified as separate objects at the rados level, distinguished by a 17 digit numeric id appended to the original, source, object id). 
radosstriper is the basis of Ponce's Ceph plugin for Scalla/XRootD\cite{poncexrootd, poncexrootd2}, which was again originally intended for use with CASTOR, but has now become a tool of wider interest, as interest in Ceph itself has grown. The Ponce XRootD-Ceph plugin, as it provides direct object access, has been adopted for testing as part of the testing work at RAL.

\section{Work at RAL}

At the UK Tier 1 at Rutherford Appleton Laboratory work is underway investigating replacement of the existing disk-only storage solution based on CASTOR with an Object Storage system, specifically with Ceph.  Running an Object Store has some cost implications for sites hosting multi petabytes of data.  Most Object Stores provide data security through replication, typically holding three or more copies of the data with security provided through a Paxos algorithm, relying on a quorum of results from checksumming of objects.  The cost of this replication at large science data centres is prohibitive.  However, since the Firefly release of Ceph, erasure coding of data is now supported, allowing us to provide the same reliability as a traditional file system with RAID 6.  Using erasure coding can limit some of the functionality (e.g. an object can not be updated directly) without the use of an additional cache pool, but the WLCG use case does not require this functionality.  If an object (or file) needs to be re-written within the experiment data management system, the whole object (or file) is replaced and replicated wherever needed.  This clearly fits well with the Ceph model for erasure coded stores.

At the Tier 1 we have deployed a test and development cluster and have been testing the CERN supplied XRootD interface to a Ceph system using radosstriper and erasure coding configured as 16+2 (i.e. the object is split into 16 chunks with 2 coding chunks) to replicate the resilience in the existing storage system.  In addition, as we are using radosstriper each chunk may be striped across several OSDs.  
This configuration ensures 89\% of the deployed storage is available for use; however, it may not be optimal as a large number of OSDs will be necessarily be involved in any write (or read) operation  leading to significant network overheads between cluster nodes. 

The test Ceph clusters are currently built on retired storage nodes and so are not optimal for Ceph, which recommends two 10GE ports (one client facing and one for internal cluster traffic) and 2GB of memory per OSD.  While the hardware deployed generally meets the specification for memory, there is only a single 10GE network port and all traffic will flow through this.

In the test setup, we have attempted to test three general data access scenarios and compare the results with a test CASTOR instance.  These three scenarios are:
`Copy + Read' where data is copied to a clients local file system using xrdcp and then some percentage of the local file is read sequentially using the file protocol;
`Direct sequential read' where data is accessed directly on the storage node and a percentage of the file is read starting from an offset of 0;
`Direct random read' where data is again accessed directly on the storage system but data is read in small chunks with a random offset between each read, until some percentage of the file is read.

The disk space used on the client in the first two tests was NFS mounted. The first two of these scenarios represent typical use cases with WLCG.  The third use case is more typical for disciplines such as climatology and seismology.  In all cases, the test files were 1GiB in extent, and each test was repeated 50 times from a random selection of 100 test files.  For the `Direct random read' scenario, two different block sizes were tested; 1MiB and 1KiB.  However, with both storage systems using the smaller block size, the consequent increases in the number of seek operations meant that even reading 10\% of the file took over 1 hour to complete. Hence, we do not include the 1KiB results in our graphs.
\subsection{Results}
\begin{figure}[h] 
   \begin{minipage}{0.49\linewidth}
   \includegraphics[width=\linewidth]{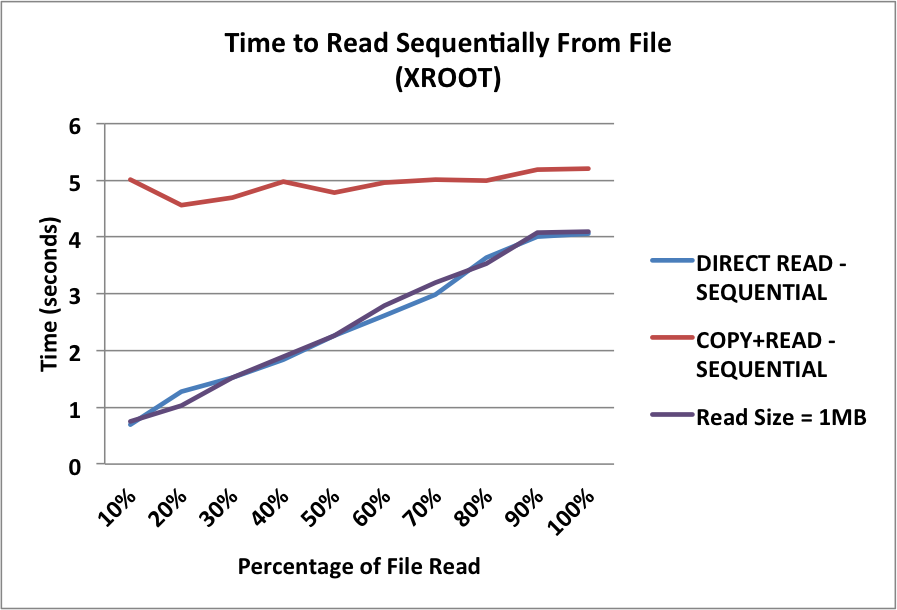} %
    \caption{\label{fig2a} Performance of file access via xrootd protocol against a CASTOR filesystem, backed by conventional storage.}
   \end{minipage}\hspace{2pc}
 \begin{minipage}{0.49\linewidth}
   \includegraphics[width=\linewidth]{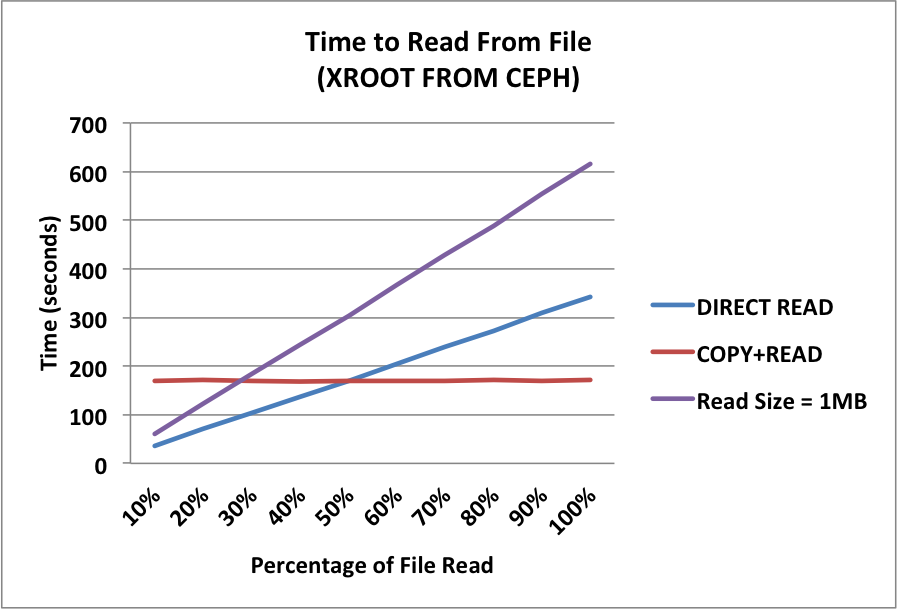} %
    \caption{\label{fig2b} Performance of file access via xrootd protocol against a Ceph Object Store, backed by erasure-coded pools.}
   \end{minipage}
\end{figure}
The results of the tests on Ceph and CASTOR are shown in figures \ref{fig2a} and \ref{fig2b}.  In the case of CASTOR, it is clear that directly reading from the storage system from a client within Rutherford Appleton Laboratory is always more performant than copying to a local disk and reading from there.  The difference between reading 100\% of the file directly and copying it represents predominantly the time write the file to disk over the NFS mount, as should be clear from the linear scaling of both of the `Direct read' graphs with percentage read.

The results from the testing on Ceph appear quite disappointing, with very poor performance in all testing. While the `Direct read' graphs do scale very linearly with percentage read, the absolute transfer time for any given fraction is over two orders of magnitude larger than that for xrootd on CASTOR. Using rados client commands on test files on a non-erasure encoded pool, we were able to retrieve the individual stripes directly and sequentially and reconstruct the file in a time comparable to that seen with CASTOR. Clearly, this indicates that the poor multiplicative performance factor for our XrootD-Ceph tests is not due to the underlying rados layer in itself. 
It was difficult to separate the potential remaining factors - the erasure coding implementation in Ceph, and the XrootD-Ceph plugin itself - as direct transfer tests were not performed on the erasure coded pool at the time of writing.

During editing of this paper, additional results from testing provided resolution to the performance issues seen. The core performance issues were due to limited IO buffer sizes inside the XrootD-Ceph interface, specifically concerning the XrootD asynchronous IO segment size. Increasing the segment size significantly, and explicitly enabling 10-way parallelism in the XrootD client's configuration, allowed a test system at CERN to saturate the Ceph pool's IO capacity over XrootD. Fixes related to this are present in releases of the XrootD-Ceph plugin, XrootD and Ceph itself at the time of publication.

\section{GFAL2-CEPH}
While interfaces for specific transfer protocols are useful, it is also important to provide access to our Object Store via a common interface layer. For Grid data management at the present date, this layer is the GFAL2\cite{GFAL2} library. Plugins for GFAL2 exist for a wide range of file access protocols (direct POSIX file I/O, HTTP, rfio, xrootd etc.), and so it represents a common glue layer, allowing interoperation between those mechanisms.

As the Ponce XRootD-Ceph plugin establishes a standard for addressing bare rados Object Stores, we implemented our GFAL2 plugin to support the same standard for addressing and representation.

\subsection{Standards Compliance}
While compatibility with emerging standards is important, it is also equally important to correctly support the wider established standards for interfaces in a field. For data access and location, these are the RFCs covering URI and URL design\cite{rfc,rfc2}. 
The pseudoURI used by Ponce's XRootD-Ceph plugin is:
\begin{verbatim}
rados[s]://user@pool,stripes,stripesize,totalsize:/OBJECTNAME
\end{verbatim}
where stripes, stripesize, totalsize are bare integer values, and OBJECTNAME is any 8-bit character sequence (including multibyte sequences without nulls), including `/' characters.

From the RFCs, we can see that this is not compliant. The ``authority" component of a URI (the part between the scheme name ({\texttt rados://}) and the first slash of the ``path'' component) should not contain encoding information concerning the data being accessed - in this pseudoURI, all of the encoding information is precisely here. The encoding data is also not labelled, and the meaning of any parameter is determined positionally, despite the information not being hierarchical in nature. Worse, the ``authority'' component also fails to correctly identify the actual authoritative entity for a Ceph cluster - one or more MON services - instead using the field to store the pool name (which is strictly part of the path component).

There is also the issue that the path we provide allows the use of the `/' character, despite this having semantic value in a compliant URL (as the separator between hierarchy levels in the path). This is intended to allow the use of paths intended for filesystem-like interfaces in this protocol in a natural way, but is also misleading in this context.
For compatibility, we support this pseudoURI, despite its issues.

We also provide an alternative, RFC-compliant, URL syntax, in which the same object and access mode as above would be represented as:
\begin{verbatim}
rados[s]o://user@mon/clustername/pool/OBJECTNAME?stripes=X&stripesize=Y&totalsize=Z
\end{verbatim}
Here, the path component correctly identifies a hierarchical namespace, beginning with the name of the host cluster and pool; the attributes of the access mode are moved into attribute-type named query strings at the end of the URL. We note that this format also makes it trivial to perform storage operations addressing multiple Ceph installs (although we also admit that the likelihood of such a requirement in production is low).

For ease of parsing, we provide four scheme names to distinguish the low-level encoding (rados v radosstriper) and schema (pseudoURI v URL).
\begin{table}[h]
\begin{center}
\begin{tabular}{|c||c|c|}
\hline
& rados & radosstriper \\
\hline
\hline
pseudoURI & rados:// & radoss:// \\
URL & radoso:// & radosso:// \\
\hline
\end{tabular}
\end{center}
\label{default}
\end{table}%

\subsection{Minimal Parsing Functionality}
In the previous section, we noted that an issue with the pseudoURI scheme, in particular, was that it implied the existence of a hierarchical filesystem where there is none, by supporting the use of the `/' character in the ``path'' component. The XRootD plugin does not perform any modifications to the path it receives, as it can rely on the surrounding XRootD mechanisms to canonicalise the string for it, resulting in a consistent final form.
Unfortunately, GFAL2 provides a completely transparent interface to its constituent plugins (the core GFAL2 API just considers a path to be a C string which can be pattern matched to determine which plugin to use to handle it), and so we cannot assume that the path component is canonicalised before it reaches us. We therefore implement a subset of the full URL canonicalisation procedure - mapping `.' and `..' segments in the hierarchy to null and `delete previous segment', and collapsing multiple sequential slashes to a single slash. We apply this canonicalisation to the incoming path to form the final object identifier, for both the pseudoURI and URL addressing schemes, as this preserves the expected functionality of the path segment of a URI. 

This approach may be interpreted as providing the weakest possible ``filesystem like" representation, in that we generate a final object identifier via an inherently hierarchical interpretation of the path string, but we regard this as a reasonable compromise between the risk of misleading the user and the risk of surprising them.

Of course, ideally, while we support all characters in object names, characters with special meaning in the path component should be encoded to suppress their effect when necessary. To this end, we also support and implement the standard percent-encoding scheme for URLs for all paths interpreted via our URL schema. (The pseudoURI schema does not support this mechanism, as it is necessary that it maintain backwards compatibility with the XRootD plugin.)

\subsection{Minimal Operations}
In order to support the functionality of the GFAL2 tools, a GFAL2 plugin needs to provide a certain minimal subset of core operations. For the most part, these are standard data access operations - {\it read} and {\it write}, sequential and random-access, `file creation' and deletion - which map trivially onto Object Storage. 
More difficult are directory and metadata operations. We follow Ponce's tool in implementing directory access at the pool level only - a pool is considered `directory-like' in that we can return a list of its contents (the flat object namespace) - and return NOENT as an error code for all other paths (even those with trailing slashes). Consistently, we also reject attempts to create directories with the EACCES error code (strictly, this is a lie, as the issue is not precisely a lack of access permissions, but there is no good way to represent topological restrictions as POSIX error codes).
Filesystem-like metadata is handled in a blended way; we fake basic file permissions metadata with a pro-forma record implying 0666 file mode, but explicitly implement xattrs as rados key-value object metadata pairs. Attempts to modify file permissions are met with EPERM errors, which again is the best fit of a bad selection of return values. 

Symlinks are explicitly forbidden for topological reasons, but, unlike the above functionality, an implementation for them is not necessary to support the GFAL2 toolset. We therefore do not implement the functions.
Checksum operations are similarly represented as (not implemented) NULL pointers at this point, as implementation is only required for a single GFAL2 tool.

With this subset of operations, the complete set of basic GFAL2 tools (except for checksumming) are supported against an arbitrary Ceph instance, at the rados and radosstriper level. Because of the simplicity of radosstriper's object striping, it is even possible to mix the two levels of operation seamlessly, creating a striped object using a {\texttt radoss://} URI, and then accessing individual chunks of the resulting stripe via {\texttt rados://} URIs.

In the example below, we can see this in practice, as we copy the passwd file in radosstriper mode, and then use rados mode to reveal the underlying chunk name (there is only one due to the small size of the file). Deleting the striped object by its bare name only succeeds with the radosstriper protocol, as that name can't be derived by rados itself.
\begin{verbatim}
> gfal-copy file:///etc/passwd radoss://admin@diamond-data:/passwdtest
Copying 1     0s  File size: 1KB
Copying 1   [DONE]  after 0s    
> gfal-ls rados://admin@diamond-data:/ | grep passwdtest
passwdtest.0000000000000000
> gfal-rm rados://admin@diamond-data:/passwdtest
rados://admin@diamond-data:/passwdtest  MISSING
> gfal-rm radoss://admin@diamond-data:/passwdtest
radoss://admin@diamond-data:/passwdtest         DELETED
\end{verbatim}

\section{Further Work}
The work reported in this paper is at a relatively early stage, and so there are many topics which remain to be developed and explored. We pick out a few aspects in this section for particular attention.

\subsection{Auth'n and Auth'z}
As mentioned above, the rados API provides a relatively simple Object Store interface. This has implications for the security and authentication aspects of storage, as well as for the aspects mentioned otherwise. Ceph's internal authorisation mechanism, cephx, provides the means to limit and control access to individual object pools (allowing read, write, execute permissions to be set) for a given user, does not, as a design principle, provide object-level restrictions. For XRootD interfaces, this does not necessarily present a problem, as XRootD also provides an X509 authentication layer which could be repurposed. However, GFAL2 does not inherently provide a security layer, so work needed to add security would involve building a wholly-novel user mapping layer, adding complexity to the interface.

\subsection{Checksumming}
At present, checksums are the only component of the GFAL2 plugin API compatible with an Object Store remaining unimplemented. There is no particular difficulty with providing a checksum interface via the rados/radosstriper APIs, as we can simply stream the object through the relevant checksumming algorithm.
However, as Ceph already generates object checksums for internal use, we believe that a more elegant solution would be to access these checksums in the first instance, avoiding needing to recalculate them in all cases. (Of course, if the requested checksum was of a different type, such as Adler32, we would still need to generate it on the fly.)

\subsection{Smart object decomposition}
At present, the interfaces we present perform a straightforward, and content-agnostic, mapping of individual files to individual objects (or equally sized objects partitioning the file). It has been suggested by others \cite{graeme} that the true benefit of Object Storage for HEP might be in the decomposition of incoming data into objects on a content-aware basis. This would presumably result in the storage of events, or other low level components of the typical ROOT file, as individual objects, relying on a mapping to provide direct access to them on demand. This is highly speculative at the time of writing, but the authors are considering if such an interface would be a feasible addition to the work.

\subsection*{}
We hope to be able to report on further work on the general field of Object Stores as physics backing at a later date, touching on some of these developments, as well as wider work at Rutherford Appleton Laboratories.

\end{document}